# Influence of portal vein occlusion on portal flow and liver elasticity in an animal model

Simon Chatelin[1] | Raoul Pop[2,3] | Céline Giraudeau[2] | Khalid Ambarki[4] | Ning Jin[5] | François Severac[1,6] | Elodie Breton[1] | Jonathan Vappou[1]

[1] ICube, CNRS UMR 7357, University of Strasbourg, Strasbourg, France;
[2] IHU-Strasbourg, Institute for Image-Guided Surgery, Strasbourg, France;
[3] Interventional Neuroradiology Department, University Hospital of Strasbourg, Strasbourg, France;
[4] Siemens Healthcare SAS, Saint Denis, France;
[5] Siemens Medical Solutions USA, Inc., Chicago, Illinois, United States;
[6] Public Healthcare Department, University Hospitals Strasbourg, Strasbourg, France.

**Correspondence**
Simon Chatelin, PhD, ICube, CNRS UMR 7357, University of Strasbourg, Bat. IHU, 1 place de l'hôpital, 67091 Strasbourg Cedex, France
Email: schatelin@unistra.fr.

**Funding information**
This work is funded by French state funds managed by the ANR (*Agence Nationale de la Recherche*) within the *Investissements d'Avenir* program for the IHU Strasbourg (Institute of Image Guided Surgery, ANR-10-IAHU-02).

**Data availability statement** Data available on request from the authors.
**Word count:** 5538  **Pages:** 15

**ABBREVIATIONS**

| | |
|---|---|
| FOV | Field of view |
| GRE | Gradient recalled echo |
| MRE | Magnetic resonance elastography |
| MRI | Magnetic resonance imaging |
| PF | Peak flow |
| PVM | Peak velocity magnitude |
| PVT | Portal venous thrombosis |
| wCV | Within-subject coefficient of variation |
| WIP | Work in progress |

**ABSTRACT**

Hepatic fibrosis causes an increase in the liver stiffness, a parameter measured by elastography and widely used as diagnosis method. The concomitant presence of portal vein thrombosis (PVT) implies a change in hepatic portal inflow that could also affect liver elasticity. The main objective of this study is to determine the extent to which the presence of portal occlusion can affect the mechanical properties of the liver and potentially lead to misdiagnosis of fibrosis and hepatic cirrhosis by elastography. Portal vein occlusion was generated by insertion and inflation of a balloon catheter in the portal vein of 4 swines. The portal flow parameters peak flow (PF) and peak velocity magnitude (PVM) and liver mechanical properties (shear modulus) were then investigated using 4D-flow MRI and MR Elastography (MRE), respectively, for progressive obstructions of the portal vein. Experimental results indicate that the reduction of the intrahepatic venous blood flow (PF/PVM decrease of 29.3%/8.5%, 51.0%/32.3% and 83.3%/53.6%, respectively) measured with 50%, 80% and 100% obstruction of the portal vein section results in a decrease of the liver stiffness by 0.8 ± 0.1%, 7.7 ± 0.4% and 12.3 ± 0.9%, respectively. While this vascular mechanism does not have sufficient influence on the elasticity of the liver to modify the diagnosis of severe fibrosis or cirrhosis (F4 METAVIR grade), it may be sufficient to attenuate the increase in stiffness due to moderate fibrosis (F2-F3 METAVIR grades) and consequently to lead to false-negative diagnosis with elastography in the presence of PVT.

## 1 | INTRODUCTION

Portal vein thrombosis (PVT) is a partial or complete occlusion by an intraluminal thrombus of the blood flow within the portal vein. The lifetime risk of developing PVT has been estimated to be around 1%[1]. The main consequence of PVT is portal hypertension and its associated complications[2]. PVT has local origin in 70% of the cases (with abdominal cancers, focal inflammatory lesions, injury to the portal venous system and cirrhosis as most frequent risk factors) and systemic origins in 30% of the cases (congenital or acquired deficiencies), and it is a common complication in patients with liver disease[3].

High prevalence of PVT is observed in the cirrhotic population[4], increasing with the severity of cirrhosis from 1% in well-compensated cirrhosis to 7.5-16% in advanced cirrhosis[5–8] and 35% in cirrhotic patients with hepatocellular carcinoma[9]. Reduction in the portal venous flow velocity is to date one of the most important predictive variables for PVT outcome in patients with cirrhosis [10]. A link between coagulation activity and progressive liver fibrosis has also been shown, most likely due to a reaction to tissue ischemic damage [11].

Liver elasticity measurement using elastography is today one of the most pertinent predictive factor as an alternative to liver biopsy allowing to estimate the fibrosis stage and presence of cirrhosis [12–17]. The liver stiffens globally as the fibrotic and cirrhotic stage increases. This stiffening is mainly due to modifications of the hepatic tissue, namely by the development of regenerative nodules surrounded by fibrous bands. Despite good agreement between elasticity measurement and fibrosis stage, a recent case study has reported portal vein thrombosis as a potential source of false-negative diagnosis of fibrosis using elastography [18]. The authors suggest that a decrease of the stiffness due to PVT could have compensated the stiffening usually observed in fibrotic livers. Recent reports of a decreased liver stiffness caused by reduced intrahepatic venous blood flow would reinforce this assumption [19,20]. Thus, the modification induced in the liver mechanical properties by fibrosis, mainly reflecting the alteration in the constitutive hepatic tissue, may be attenuated by portal flow modifications in the presence of PVT.

In the present study, we proposed an original preclinical investigation of the consequences of intraluminal portal vein occlusion on both portal flow and liver elasticity, in order to answer the following question: to what extent is the presence of PVT likely to influence liver elasticity, and thereby potentially degrade the diagnosis of liver fibrosis or cirrhosis by elastography? In this study, modifications of the venous portal inflow were obtained *in vivo* in 4 animal subjects (swine) through various levels of portal vein obstruction with a balloon catheter. For each level of portal vein obstruction, portal vein flow and liver elasticity were quantified non-invasively using 4D-flow MRI and MR-elastography (MRE), respectively. The *in vivo*, non-invasive quantification of the vascular and mechanical effects of portal obstruction on the liver are thus obtained.

## 2 | MATERIALS AND METHODS

The study was conducted on 4 female swines, weighting respectively 27, 25, 27 and 30 kg for subjects # 1 to #4. Anesthesia was induced using intravenous Zoletil (20cc, *Virbac*) and was maintained using 2% isoflurane (*Forene, AbbVie*). All experiments were performed in agreement with the European Community Council Directive (010/63/UE) and in accordance with national animal experiment regulations (APAFIS #14092-2018031513247711 v1). A surgical procedure was first performed under angio-scanner monitoring in order to introduce the balloon catheter for portal venous flow obstruction, followed by the MRI protocol.

## 2.1 | Control of the portal flow

Occlusion of the portal vein was obtained using a balloon catheter (*XXL<sup>TM</sup> Vascular, Boston Scientific*), with dimensions adapted to the portal vein diameter, measured with X-ray fluoroscopy. Percutaneous access to an intrahepatic branch of the portal vein was performed under ultrasound (*Acuson S3000, Siemens Healthcare*) and X-ray fluoroscopy guidance (*Artis Zeego* system, *Siemens Healthcare*). The balloon catheter was then introduced in the main trunk of the portal vein under fluoroscopy guidance. The inflation volume in the balloon was calibrated under fluoroscopy for each pre-determined degree of portal occlusion (0%, 50%, 80% and 100% of the portal vein diameter). For that purpose, a 1:1 iodine contrast (*Visipaque 270, GE Healthcare*) in normal saline solution was used for balloon inflation. Representative angiography images for all 4 animals are reported in figure 1(B), for a balloon catheter dilatation of 80% of the intraluminal portal vein section.

## 2.2 | MRI procedure

MRI acquisitions were conducted in a 1.5T MRI (*MAGNETOM Aera, Siemens Healthcare*) using the in-bore body coil, with the animal lying in supine position. Normal saline was used for balloon inflation. The inflation volume in the balloon catheter was varied in order to obtain successive intraluminal obstructions of approximately 0% (reference, no inflation), 80%, 50%, 0% (control) and 100% of the portal vein section. Actual obstruction levels were evaluated after the procedure in post-treatment of the SPIRAL-VIBE 3D MRI acquisitions. This experimental order makes it possible not only to avoid total occlusion, potentially risky for the animal, from the beginning of the procedure, but also to have a control point in the middle of the procedure, given that anesthesia or previous occlusions may have induced physiological effects leading to a variation in liver elasticity. For each inflation level of the balloon catheter, the following MRI sequences were successively acquired (according to the experimental protocol illustrated in figure 1(A)):
1. LOCA: fast scout localizer images.
2. T1-VIBE: for precise positioning of the field of view (FOV) of the following sequences.
3. MRE #1: these acquisitions aim at mapping and quantifying the liver elasticity. The

MRE acquisitions were performed in the liver using a gradient recalled echo (GRE) sequence with flow-compensated motion sensitizing gradients and mechanical vibration at 60 Hz using a pneumatic exciter (*Resoundant®*).

4. 4D-flow: the 4D-flow sequence aims at investigating over time in 3D the peak flow (PF) and the peak velocity magnitude (PVM) in the portal vein and vena cava. Acquisitions were obtained using the research prototype 4D flow pulse sequence [21] WIP785A (*Siemens Healthcare*). Flow parameters were quantified using the prototype "4DFlow v2.4" WIP postprocessing software (*Siemens Healthcare*). Phase anti-aliasing and motion tracking filtering were applied as post-treatment before flow quantifications. PF and PVM are estimated downstream of the balloon in the main trunk of the portal vein, halfway between the end of the balloon and the division into right and left portal branches.
5. MRE #2: repeat.
6. SPIRAL-VIBE: the high-resolution images aim at measuring the diameter of both the balloon catheter and the portal vein. Acquisitions were performed using UTE spiral 3D-GRE sequence with volumetric interpolation (VIBE). The intraluminal sectional vein obstruction was then calculated for each of the theoretical inflation state as the ratio of the balloon diameter to the portal vein diameter, measured in the SPIRAL-VIBE images.

Images were acquired in the coronal plane (except for MRE images acquired in an oblique plane aligned against the inclination of the pneumatic vibrator, i.e. MRE images acquired perpendicular to the active face of the pneumatic vibrator). Main acquisition parameters are summarized in Table 1. Animal was under breath-hold (mechanical ventilation stopped) during T1-VIBE and MRE acquisitions. The 4D-flow sequence was respiratory-gated using a pressure belt sensor. Because of the long acquisition time of the 4D-flow, the MRE acquisition was performed before and after the 4D-flow acquisition. The MRE shear modulus used in the analysis corresponds to the average of these 2 measurements. Representative angiography, 4D-flow, MRE waves, MRE stiffness maps and SPIRAL-VIBE MRI images are shown for all 4 animals in figure 1(B).

| Parameter | T1-VIBE | MRE 60 Hz | 4D-flow | SPIRAL-VIBE |
|---|---|---|---|---|
| TR/TE [ms] | 4.72/2.18 | 50/23.75 | 51.68/3.77 | 4.57/0.05 |
| TT [ms] | - | - | 20 to 440 by 52.5 steps (9 frames) | - |
| Flip angle [°] | 10 | 25 | 8 | 5 |
| FOV [mm x mm] | 243 x 300 | 239 x 300 | 284 x 414 | 300 x 300 |
| Matrix [-] | 156 x 256 | 204 x 256 reco pas acq | 106 x 192 | 288 x 288 |
| Number of slices [-] | 72 | 3 | 20 | 176 |
| Slice thickness [mm] | 2 | 5 | 2.5 | 1 |
| Bandwidth frequency [Hz/pixel] | 350 | 260 | 450 | - |
| Number of time frames [-] | - | - | 11 | |
| Encoding velocity [cm/s] | - | - | 50 (in all directions) | - |
| Duration [s] | 18 | 67 | ~518 | ~278 |
| Breathing [-] | Hold | Hold | Free - respiratory-gated | Free |

**TABLE 1**   Relevant acquisition parameters for the MRI sequences.

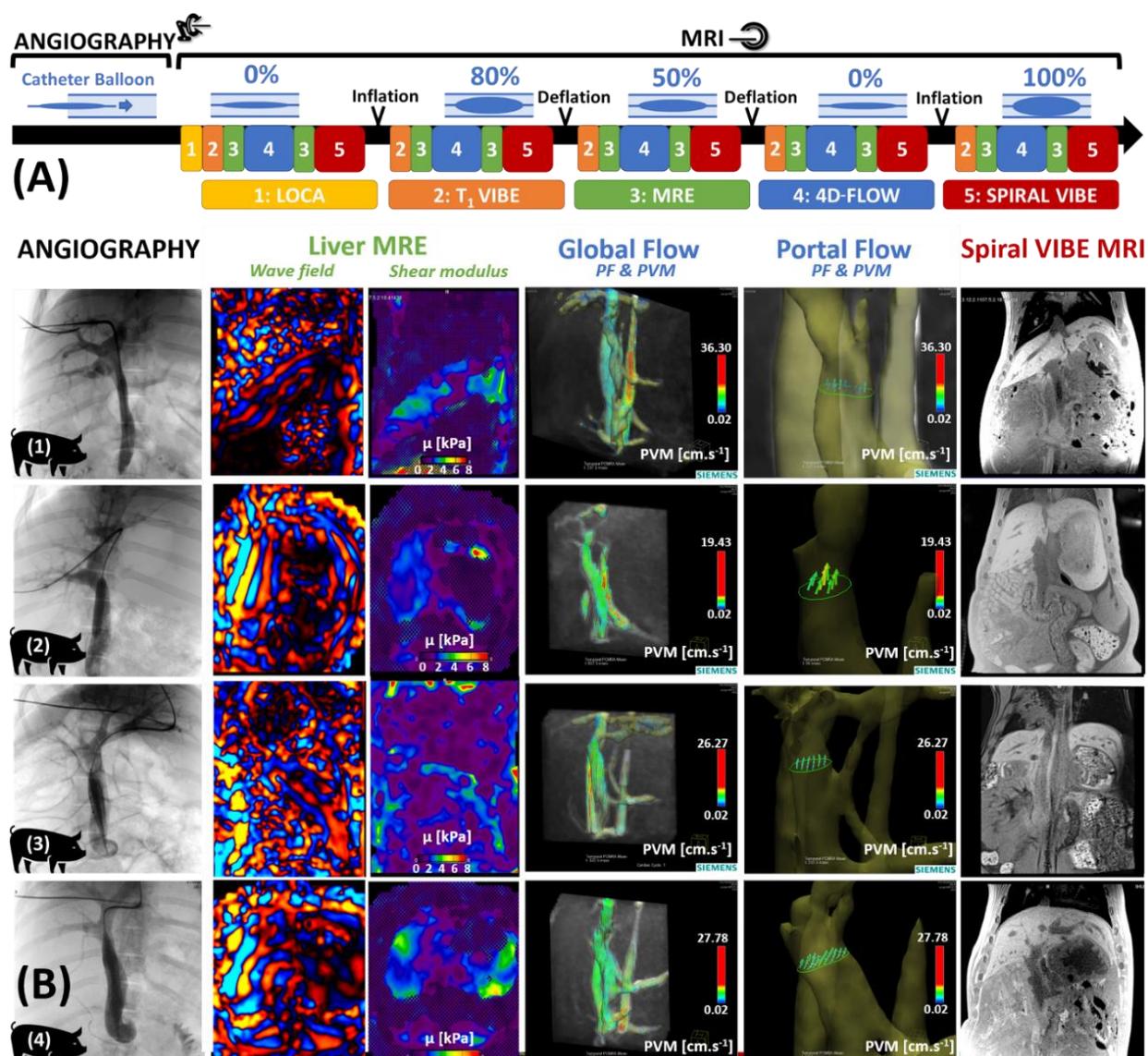

**FIGURE 1**   Timeline of the experimental protocol (A). After positioning of the catheter through the portal vein under X-ray guidance, MRI acquisitions alternating 4D-flow, MRE and anatomical acquisitions were performed for successive dilation states of the balloon. Representative angiography, 4D-flow, MRE wave, MRE stiffness map and SPIRAL-VIBE MRI images are presented for all 4 animal models (B).

## 2.3 | Analysis

The liver stiffness and flow values were expressed as mean ± standard deviation.

The statistical repeatability of MRE measurements was evaluated using R Core Team 2019 v3.6 [22]. The within-subject coefficient of variation (wCV) was evaluated with 95% lower limit and upper limit confidence intervals, including variance stabilizing transformation, [23,24] and interpreted based on the criteria defined in [25].

The individual correlation between the liver stiffness and the portal PVM or PF was calculated for each animal using Spearman's $r_s$ rank correlation coefficient in order to investigate potential monotonic relationships (statistical analysis performed using the XLSTAT add-in, Addinsoft, Paris, France [26]).

## 3 | RESULTS

### 3.1 | Occlusion of the portal vein and flow quantification

In addition to nominal balloon dimensions (diameter and length), the intraluminal sectional vein obstruction, balloon diameter and portal vein diameter measured by MRI are reported in Table 2. The measured sectional obstructions were close to the expected values (with a maximal difference of 3.3%). Dextrocardia was observed in animal subject #3 (figure 1(B)).

Baseline PVM and PF were measured at initial 0% occlusion as 15.53 ± 2.45 cm.s$^{-1}$ and 13.46 ± 1.63 mL.s$^{-1}$, respectively, averaged over subjects. The individual evolutions of both the portal PF and PVM obtained from 4D-flow data are shown in figure 2(A) (orange and blue curves, respectively) against the sectional obstruction level of the portal vein. Note that for readability, the scale for subject #3 is different from the one for other subjects. It should be noted that for 100% occlusion, defined as the balloon diameter being equal to the intraluminal portal vein section, portal vein compliance (elasticity) ensures remnant portal flow. For nominal intraluminal occlusions of the portal vein of 50%, 80% and 100%, the mean decrease of the portal PF was 29.3%, 51.0% and 83.3%, respectively, compared to the initial 0% occlusion reference. Similarly, the portal PVM was reduced by 8.5%, 32.3% and 53.6% for nominal intraluminal occlusions of 50%, 80% and 100% of the portal vein, respectively, compared to the initial 0% occlusion reference PVM. Compared to the second "control" 0% occlusion state, the mean decrease in portal PF and PV with 100% portal vein obstruction were 82.2% and 52.6%, respectively, i.e. similar to the ones found compared to the initial 0% occlusion state. Hence the flow variations found when using the first "initial" 0% occlusion or the second "control" 0% occlusion were similar.

| Animal # | Parameters | Nominal intraluminal sectional vein occlusion [%] | | | | Nominal balloon diameter / length [mm] |
|---|---|---|---|---|---|---|
| | | **0%** | **50%** | **80%** | **100%** | |
| 1 | Balloon diameter [mm] | 0.6 | 5.8 | 7.9 | 9.1 | 9 / 40 |
| | Portal vein diameter [mm] | 8.7 | 7.9 | 8.9 | 9.1 | |
| | **Sectional obstruction [%]** | **0.4** | **52.5** | **79.8** | **100.0** | |
| 2 | Balloon diameter [mm] | 1.0 | 8.3 | 10.2 | 12.3 | 12 / 40 |
| | Portal vein diameter [mm] | 11.5 | 11.8 | 11.5 | 12.3 | |
| | **Sectional obstruction [%]** | **0.8** | **50.1** | **79.4** | **100.0** | |
| 3 | Balloon diameter [mm] | 2.8 | 12.7 | 15.6 | 18.2 | 18 / 40 |
| | Portal vein diameter [mm] | 17.9 | 17.6 | 17.6 | 18.2 | |
| | **Sectional obstruction [%]** | **2.4** | **52.1** | **78.6** | **100.0** | |
| 4 | Balloon diameter [mm] | 0.2 | 11.7 | 14.7 | 16.8 | 18 / 40 |
| | Portal vein diameter [mm] | 16.0 | 16.2 | 16.1 | 16.8 | |
| | **Sectional obstruction [%]** | **0** | **52.2** | **83.3** | **100.0** | |

**TABLE 2** Individual intraluminal sectional vein occlusions as determined on the SPIRAL-VIBE images

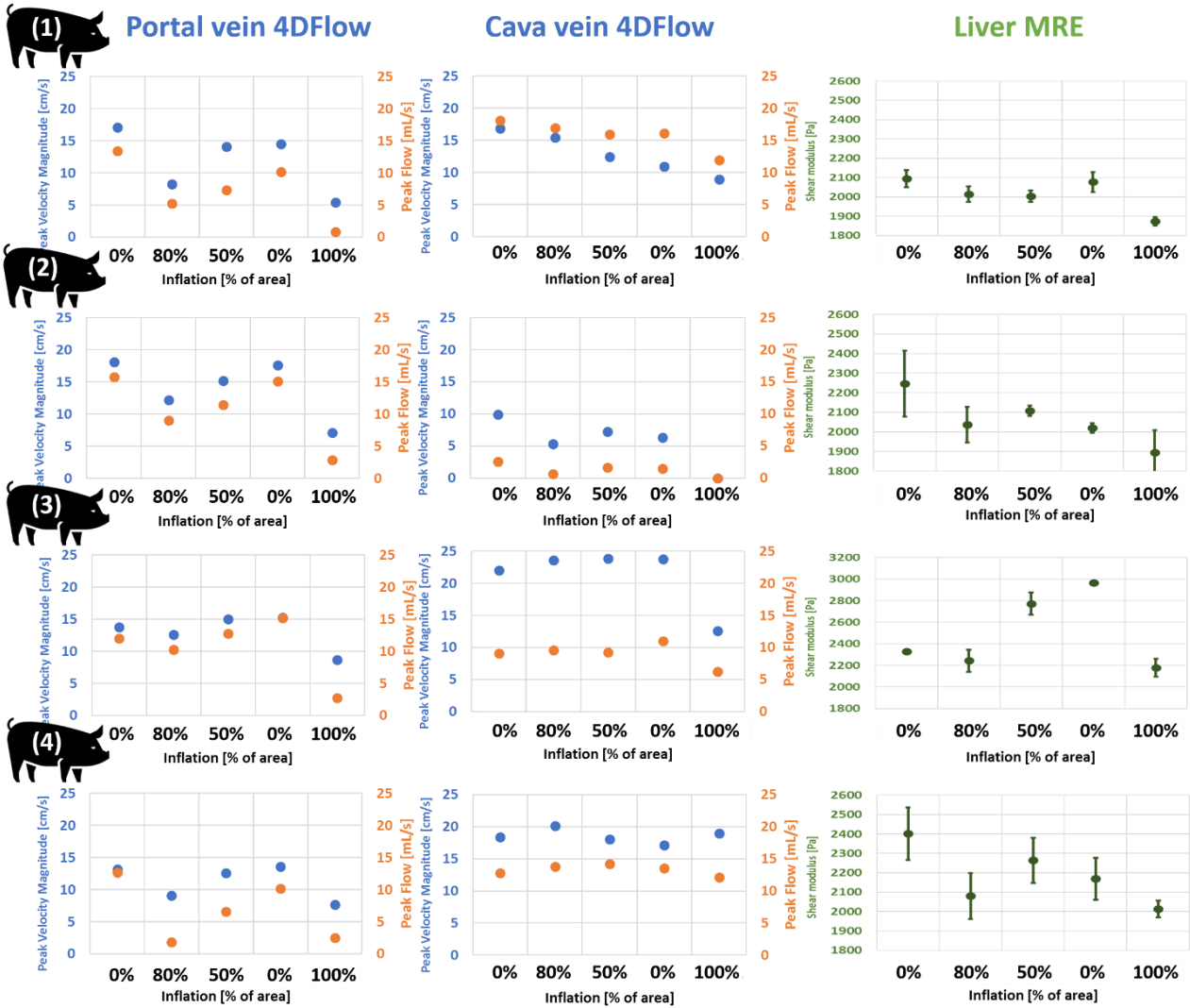

**Figure 2.** Individual evolution of the flow parameters PVM (blue curves) and PF (orange curves) in both the portal (left column) and vena cava (central column) as a function of the portal vein obstruction. Similar evolution was found for the liver stiffness (right column) where the standard deviation bars correspond to variability over the pooled data from the two MRE scans. Note that the shear modulus scale is different in subject #3 compared to other subjects.

PF and PVM were also quantified in the vena cava. These values did not seem to be influenced by the portal occlusion, except for 100% occlusion measurements for subjects #2 and #3. While the flow parameters were stable in the vena cava for subjects #3 and #4, their values slowly and regularly decreased for subjects #1 and #2.

## 3.2 | Influence on the liver elasticity

A good statistical repeatability between the repeated MRE stiffness measurements was observed for the initial 0%, the 80%, the 50%, the second 0% and the 100% inflation states, with wCV values (with 95% lower limit and upper limit confidence intervals) of 4.87% (2.43%; 9.78%), 4.44% (2.22%; 8.89%), 3.49% (1.74%; 7.16%), 2.67% (1.32%; 5.56%) and 3.72% (1.86%; 7.47%), respectively. In general, the repeatability of the MRE measurement was good, with a wCV general value of 3.90% (2.85%; 5.85%). With wCV values inferior to 7%, these observations are in accordance with the RSNA-QIBA consensus [25] and tend to reinforce the confidence in the stiffness values, despite the small number of subjects.

Baseline liver stiffness was measured at initial 0% occlusion as 2,268 ± 131 Pa. Individual evolutions of the liver stiffness are shown in figure 2 (green curves) against the sectional obstruction of the portal vein. The global evolution of the liver's shear modulus was shown as a function of the portal vein obstruction for the whole cohort as box plots in figure 3(A). Mean liver softening of 0.8 ± 0.1% and 7.7 ± 0.4% were observed for nominal occlusions of 50% and 80% of the portal vein, respectively, compared to the initial 0% occlusion. A mean variation of 12.3 ± 0.9% (compared to the initial 0% inflation) to 13.8 ± 1.0% (compared to the second 0% inflation) was observed for the 100% obstruction of the portal vein.

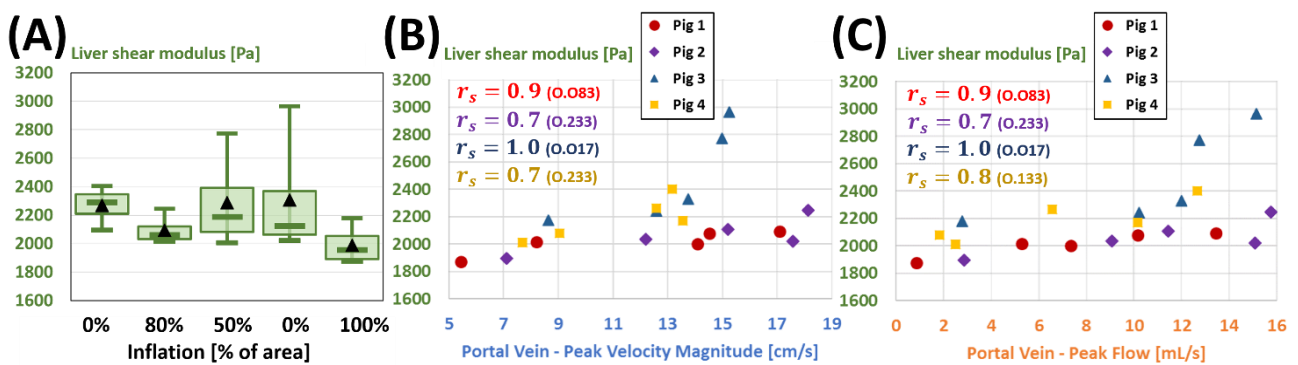

**Figure 3.** Box plots representing the evolution of liver stiffness for the whole cohort with the successive degrees of intraluminal portal vein occlusion (A). The liver stiffness was shown as a function of the portal PVM (B) and PF (C) for the four subjects. Spearman's $r_s$ rank correlation coefficients close to 1 indicate individual monotonic relationships. The level of significance of the correlation is indicated as its p-value in the parenthesis.

Individual liver elasticities were plotted as a function of the portal PVM (figure 3(B)) and PF (figure 3(C)) for each nominal portal vein occlusion. Significant Spearman's $r_s$ rank correlation was observed between the liver mechanical and portal flow parameters, with the possibility of identifying linear relationships, as attested by the correlation coefficients close to 1. This relationship was observed for each animal individually, but the limited number of subjects combined with inter-individual variations did not allow concluding on a significant general relationship between liver mechanical property and portal flow.

## 4 | DISCUSSION

This study investigated how liver elasticity can be affected by portal vein obstruction and by the corresponding inflow modifications in an animal model. Previous clinical studies showed that a reduction of the intrahepatic venous blood flow due to PVT results in a significant decrease of the liver stiffness [19,20]. For instance, a decrease of the liver stiffness of 38% has been associated to intrahepatic portal flow decrease of 30% in the context of the Valsalva maneuver [20]. In the present pre-clinical study, the decrease in portal flow was induced by several levels of occlusion of the intraluminal section of the portal vein. The flow parameters were quantified in an original way using 4D-flow MRI with no portal vein occlusion and with three levels of portal vein section occlusion (50%, 80% and 100%). The important decreases observed in PF and PVM with portal occlusions were associated with limited decreases in average liver stiffness of 0.8%, 7.7% and 12.3%, respectively for 50%, 80% and 100% portal vein occlusions.

This study has been conducted in swines. The external morphology of the porcine liver is well documented [27], however, few studies have detailed the intraparenchymal vascular anatomy compared to humans. Anatomically, the vascular system and the portal system of swine have been found to be similar to the human ones [28]. The baseline flow values measured in this study in swines were lower (15.53 ± 2.45 cm.s$^{-1}$ and 13.46 ± 1.63 mL.s$^{-1}$ for PF and PVM, respectively) but in the same range of magnitude than those reported in humans in the literature, with PF of 18.3 ± 6.7 mL.s$^{-1}$ [29] and 19.5 ± 6.7 mL.s$^{-1}$ [20], and PVM of 22.14 ± 4.87 cm.s$^{-1}$ [20]. Hence anatomical and portal flow similarities tend to indicate that swine is a valid animal model for vascular effects of PVT on liver elasticity.

The increase of liver stiffness with fibrosis results from a well-known "histological" effect, i.e. from modifications in the constitutive hepatic tissue [30]. For instance, an average increase in liver stiffness of 14% was reported in a previous study [17] using MRE between patients without (F0–F1 grades, according to the METAVIR scoring system) and with substantial moderate (F2–F3) fibrosis, while this increase reached 109% between patients without (F0–F1 grades) and with advanced fibrosis or cirrhosis (F4). In the presence of PVT concomitant with hepatic fibrosis, changes in intrahepatic portal flows may occur due to partial or total occlusion of the portal system. Thus, in accordance with the vascular effects described in this study, these flow changes would cause a decrease in liver stiffness that may attenuate the "histologically" induced increase in stiffness. Given the order of magnitude of the decrease in stiffness caused by PVT-related vascular effects compared to the increase observed with "histological" effect [16,17], portal vascular occlusion is unlikely to have any adverse effect on the diagnosis of elevated fibrosis stages and hepatic cirrhosis by elastography. This confirms the good correlation observed between high fibrosis stage and stiffness measurement [31]. However, the order of magnitude of the clinical "histological" increase in stiffness reported with moderate stages of fibrosis is comparable to the one of the "vascular" decrease in stiffness found in this study in swines. Although the occurrence of concomitant PVT at moderate stages of fibrosis remains low [32], there is still a risk of false-negative results in the diagnosis of moderate fibrosis, as recently suggested in a clinical case study [18]. This tends to suggest the need to couple elastography measurements with PVT evaluation in the detection of early fibrosis based on liver mechanical properties.

The quantifications of PF and PVM in the vena cava do not indicate any influence of the portal occlusion, except for 100% occlusion measurements for the animal subjects #2 and #3. For these two subjects, 100% portal occlusion appeared to have systemic vascular implications, but no conclusion can be drawn out of the limited number of subjects. While the flow parameters were stable in the vena cava for two animal subjects (#3 and #4), their values slowly and regularly decreased for the two other subjects (#1 and #2), suggesting a potential influence of experimental conditions, such as anesthesia [33,34] and previous portal vein occlusions. Overall, the systemic vascular variations revealed in the vena cava remain much smaller than those observed in the portal vein with intraluminal occlusions, and their influence on portal flow variations was hence neglected in this work.

It must be noted that subject #3 gave signs of reawakening within 10 minutes of the 100% occlusion, including motion artifacts in MR images. Hence the isoflurane level was increased, and all acquisitions were repeated for 100% occlusion once the anesthesia was deep enough. The fact that the anesthesia may have been insufficiently deep could explain why the liver elasticity and flow values measured in this subject for the 50% and the second 0% occlusion levels, are substantially different from those found in the other subjects. However, no motion artifact was found in the MR images for neither 50% nor second 0% occlusion levels. In the absence of formal evidence of insufficient anesthesia depth for these 2 levels, we considered that this subject should not be excluded from this study, but its values should be considered with care.

Opposite to the increase observed for subject #3, the decrease observed in liver stiffness for subjects #1, #2 and #4 between the initial and "control" 0% occlusion levels may be attributed to experimental biases such as the vascular effect of lengthened anesthesia and remnant effects from the two occlusion levels (80% and 50%) induced before the second "control" 0% occlusion level.

This preliminary study was conducted in a limited number of subjects. This preliminary study was conducted in a limited number of subjects. Please note that, in the absence of prior exclusion criteria, we decided to maintain the data for animal #3 despite the presence of dextrocardia. A large inter-individual variability in the size of the portal vein (diameters from 8.7 to 17.9 mm for animals with similar weight) has been observed. Hence the nominal balloon diameter was individually adapted in order to obtain pre-defined sectional occlusion levels of the portal vein. Despite the limited number of subjects, satisfying repeatability of the MRE measurements were found for each occlusion levels according to the RSNA-QIBA consensus. However, inter-individual variations and variations between the two 0% occlusion (i.e. initial and "control") are important and limit the conclusions to be drawn out of this preliminary study. Variations between the two 0% occlusion levels highlight the presence of experimental biases, which influence on the sensitivity of elasticity measurements resulted in no change found for the 50% occlusion while an 8 to 12% change in elasticity was found for 80 and 100% occlusion levels. Additional work in larger cohort is required in order to reinforce the conclusions concerning the effect of partial obstruction of the portal vein on flow and stiffness measurements. Future work is required in order to reinforce the conclusions concerning the effect of partial obstruction of the portal vein on flow and stiffness measurements. Please note that, in the absence of prior exclusion criteria, we decided to maintain the data for animal #3 despite the presence of dextrocardia. A large inter-individual variability in the size of the portal vein (diameters from 8.7 to 17.9 mm for animals with similar weight) has been observed. Hence the nominal balloon diameter was individually adapted in order to obtain pre-defined sectional occlusion levels of the portal vein. Besides the limited number of subjects, a satisfying repeatability in the evolution of both the flow and mechanical parameters with several levels of portal vein occlusion was found.

Although the evidence of a link between vascular fluid pressure and mechanical properties of the liver has been repeatedly reported [35–40], the relationship between vascular flow and hepatic mechanical properties remains poorly described and very rarely mentioned [18]. In this work, an experimental protocol was proposed to investigate the latter relationship and preliminary results were reported in a limited number of subjects. Further similar investigation is necessary with the additional measurement of portal pressure or portal resistance, in order to understand more accurately the mechanisms linking vascular flow, flow velocity and liver mechanical properties. Such investigation would help determine whether the acute obstruction may represent a typical sinusoidal regulation and the pathophysiologic response of the liver against chronic portal thrombosis. Otherwise, further investigation concerning the kinetics of flow and elasticity changes after the occlusion would be of interest as this point was not investigated in this study.

In conclusion, this pre-clinical study illustrated the underlying vascular effect of PVT on liver stiffness measured by elastography. The influence of the portal vein obstruction was observed not only on the portal flow PVM and PF with an original use of 4D-flow MRI, but also on liver stiffness using MRE. Results showed a decrease of liver's shear modulus correlated with the portal flow reduction. This study confirms that vascular effects due to the frequent presence of PVT concomitantly with an acute fibrotic stage or cirrhosis are not likely to alter the diagnosis of the latter by liver elastography. Conversely, moderate fibrosis causes only a slight increase in liver stiffness that may be attenuated in the presence of PVT by the described vascular effects, potentially leading to false negative in the diagnosis of moderate fibrosis using elastography. The vascular implications of PVT on liver elasticity reported in this study are so of great interest in optimizing the diagnosis of moderate hepatic fibrosis with elastography. Beyond hepatic fibrosis, the influence of blood flow on elastography measurements could be investigated similarly in different organs and pathologies involving both mechanical and vascular modifications, such as polycythemia, certain cancers (liver, pancreas, kidney, or adrenal gland), injury or blood clotting disorders.

**ACKNOWLEDGMENT**

The authors would like to thank Dr. Thomas Benkert (*Siemens Healthcare GmbH, Erlangen, Germany*) for providing the prototype SPIRAL VIBE sequence, the staff of the imaging platform of the IHU-Strasbourg for their support and availability during the animal experiments and the IRIS platform (ICube laboratory) for technical support.

**FUNDING INFORMATION**

This study has received funding by French state managed by the ANR (*Agence Nationale de la Recherche*) within the *Investissements d'Avenir* program for the IHU Strasbourg (Institute of Image Guided Surgery, ANR-10-IAHU-02).

**ETHICS APPROVAL**

Approval from the institutional animal care committee was obtained. All experiments were performed in agreement with the European Community Council Directive (010/63/UE) and in accordance with national animal experiment regulations (APAFIS #14092-201803151324771 v1).